\documentclass[12pt,preprint]{aastex}

\slugcomment{submitted to ApJ}

\shorttitle{H $\alpha$ absorption in a BAL quasar}
\shortauthors{Aoki et al.}

\begin{document}

\title{Discovery of H$\alpha$ absorption in the unusual broad absorption
line quasar
SDSS~J083942.11+380526.3
\footnote{Based in part on data collected at Subaru Telescope, which is
operated by the National Astronomical Observatory of Japan.}}

\author{Kentaro Aoki\altaffilmark{1}, Ikuru Iwata\altaffilmark{2}, Kouji
Ohta\altaffilmark{3}, Masataka Ando\altaffilmark{3}, Masayuki
Akiyama\altaffilmark{1}, and Naoyuki Tamura\altaffilmark{4,5}}
\altaffiltext{1}{Subaru Telescope, National Astronomical Observatory of
Japan,
    650 North A'ohoku Place, Hilo, HI 96720; kaoki, akiyama@subaru.naoj.org}

\altaffiltext{2}{Okayama Astrophysical Observatory, National
Astronomical Observatory of Japan,
Okayama 719-0232, Japan; iwata@oao.nao.ac.jp}

\altaffiltext{3}{Department of Astronomy, Kyoto University,
Kyoto 606-8502, Japan; ohta, andoh@kusastro.kyoto-u.ac.jp}

\altaffiltext{4}{Department of Physics, University of Durham, Durham DH1
3LE, U.K.}
\altaffiltext{5}{Present address: Subaru Telescope, 
National Astronomical Observatory of Japan, 650 North A'ohoku Place, Hilo, 
HI 96720; naoyuki@subaru.naoj.org}

\begin{abstract}
We discovered an H$\alpha$ absorption in a broad H$\alpha$ emission line
of an unusual broad absorption line quasar,
SDSS~J083942.11+380526.3 at $z=2.318$,
by near-infrared spectroscopy with the Cooled Infrared Spectrograph
and Camera for OHS (CISCO) on the Subaru telescope.
The Presence of non-stellar H$\alpha$ absorption is known only
in the Seyfert galaxy NGC 4151 to date, thus our discovery is the
first case for quasars.
The H$\alpha$ absorption line is blueshifted by 520 km s$^{-1}$
relative to the H$\alpha$ emission line, and its redshift almost coincides
 with
those of UV low-ionization metal absorption lines.
The width of the H$\alpha$ absorption ($\sim 340$ km s$^{-1}$) 
is similar to those of the UV low-ionization absorption lines.
These facts suggest that the H$\alpha$ and the low-ionization metal absorption
lines are produced by the same low-ionization gas
which has a substantial amount of neutral gas.
The column density of the neutral hydrogen is estimated to be $\sim
10^{18}$ cm$^{-2}$ by assuming a gas temperature of 10,000 K 
from the analysis of the curve of growth.
The continuum spectrum is reproduced by a reddened ($E(\bv) \sim 0.15$ mag 
for the SMC-like reddening law)
composite quasar spectrum.
Furthermore, the UV spectrum of
SDSS~J083942.11+380526.3 shows a remarkable similarity to that of
NGC 4151 in its low state,
suggesting the physical condition of the absorber in SDSS~J083942.11+380526.3 is
similar to that of NGC 4151 in the low state.
As proposed for NGC 4151, SDSS~J083942.11+380526.3 may be also seen through
the close direction of
the surface of the obscuring torus.
\end{abstract}


\section{INTRODUCTION}
Broad absorption line (BAL) quasars are characterized by the absorption 
troughs of UV resonance lines
of which velocity widths are typically $\sim 2,000 - 20,000$ km s$^{-1}$. 
The troughs are blueshifted relative to emission
lines from 2,000 km s$^{-1}$ to as much as $\sim$ 0.2c.
BAL quasars are divided into three subtypes
depending on what kind of ions seen as absorption.
High-ionization BAL quasars (HiBALs) show absorption
from \ion{C}{4}, \ion{N}{5},
\ion{Si}{4} and Ly$\alpha$.
Low-ionization BAL quasars (LoBALs) show
absorption from \ion{Mg}{2}, \ion{Al}{3}, and \ion{Al}{2},
in addition to the high-ionization absorption.
A fraction of LoBALs show absorption from excited fine-structure
levels of the ground term and excited terms of \ion{Fe}{2} and
\ion{Fe}{3} \citep{Haz87,Beck97, Beck00, Hal02}.
They are called iron LoBALs (FeLoBALs).
BAL quasars occupy 10-20\% of
optically selected quasars \citep{Wey91, HF03, Rei03}.
The fraction of LoBALs among BAL quasars is 13-15\%, and $\sim 2$\% in
all quasars \citep{Wey91,Rei03}.
FeLoBALs are rare and comprise only 15\% of LoBALs \citep{Hal02}.
\par
The number of known FeLoBALs has been increasing.
After discovery of the first FeLoBAL LBQS 0059-2735 \citep{Haz87},
\citet{Cow94} discovered a similar object to this in
their K-band survey (Hawaii 167).
\citet{Beck97, Beck00} also discovered five FeLoBALs in their
radio-selected quasar sample, FIRST Bright Quasar Survey
\citep[FBQS;][]{Whi00}.
The Sloan Digital Sky Survey \citep[SDSS;][]{Yor00} is dramatically
increasing the number of FeLoBALs.
\citet{Rei03} discovered 10 FeLoBALs in SDSS Early Data Release
quasar catalog \citep{Sch02}.
\citet{Hal02} identified more than one dozen of FeLoBALs whose
characteristics of spectra are much different from those of previously
known BAL quasars.
Some of them have tremendous numbers of UV absorption lines.
Others have absorption troughs which removes almost ($\sim 90$\%) all the flux
shortward of \ion{Mg}{2}.
At least one FeLoBAL has \ion{Fe}{3} absorption but does not have 
\ion{Fe}{2} absorption.
Many of them are heavily reddened by up to $E(\bv)\simeq0.5$ mag.
Such unusual FeLoBALs may occupy a new parameter space (the black hole mass, mass accretion
rate, viewing angle, outflow rate etc.) of quasars.
An Extensive search for such BAL quasars is therefore important to
understand the structure and evolution of quasars.
\par
There is still no consensus concerning whether BAL quasars are
intrinsically different from non-BAL quasars.
\citet{Wey91} pointed out the close similarity in the
emission-line and
continuum properties of BAL quasars and non-BAL quasars.
They concluded that BAL quasars are non-BAL quasars seen from a different
viewing angle, i.e., both types of quasars belong to the
same population.  
The results of spectropolarimetry of BAL quasars
\citep[e.g.,][]{GM95, Cohen95, HW95, Ogle99, SH99} 
have also been interpreted in such a way
that BAL quasars are quasars
seen through the edge of the obscuring tori.
On the other hand, some authors argue that BAL quasars, especially
LoBALs, are young or recently refueled quasars \citep{BM92, Voit93, Beck00}.
They infer that the absorption troughs of BAL quasars
appear when the nuclei are blowing the obscuring dust off;
through this BAL phase, quasars may evolve from a
dust-enshrouded stage (the infrared luminosity is
therefore expected to be large) to a normal quasar phase
\citep{San88}.
All four LoBALs/FeLoBALs currently known at $z < 0.4$ are, in fact,
ultraluminous infrared ($L_{\rm IR} > 10^{12} L_{\sun}$) galaxies and
major mergers \citep{CS02}.
The discovery of a population of unusual BAL quasars \citep{Hal02} may
suggest
there is a population in different evolutionary phase or structure. 
\par

During our search for LoBALs and FeLoBALs by visual inspection of
$\sim 4,800$ spectra between redshifts of 2.1 and 2.8 in the SDSS
Data Release 3 \citep[DR3;][]{Aba05},
we found an unusual BAL quasar,
SDSS~J083942.11+380526.3 (hereafter SDSS~J0839+3805).
Photometric data of SDSS~J0839+3805 from SDSS DR3 and the Two Micron All Sky Survey (2MASS) Point Source 
Catalog are tabulated in Table~\ref{tbl1}.
The absolute magnitude in the $i$ band of SDSS~J0839+3805 is $-26.8$ mag (AB)
with the $k-$correction
\citep{Sch05} by assuming $H_{\rm 0} = 70$ km s$^{-1}$ Mpc$^{-1}$, 
$\Omega_{m} = 0.3$, and $\Omega_{\Lambda} = 0.7$.
Figure \ref{fig1} shows its rest UV spectrum from
SDSS data.
SDSS~J0839+3805 is a FeLoBAL; it shows the absorption in 
the excited-state \ion{Fe}{2} as well as \ion{C}{4}
$\lambda\lambda 1548, 1551$ and \ion{Al}{3} $\lambda\lambda 1855, 1863$.
The troughs at wavelengths between 7700 \AA~and 8000 \AA~and between 
8500 \AA~and 8700 \AA~are from
the excited levels of the ground term and the excited term of \ion{Fe}{2}.
It also shows absorption of low-ionization metal lines at $z=2.31$ such as
\ion{Si}{2} $\lambda\lambda 1527, 1533$, \ion{Al}{2}  $\lambda 1671$,
\ion{Zn}{2} $\lambda\lambda 2026, 2063$, and
\ion{Cr}{2} $\lambda\lambda\lambda 2056, 2062, 2066$.
The spectrum is very similar to those of so-called
``many-narrow-trough'' BAL quasars,
especially SDSS~J112526.13+002901.3 \citep{Hal02}.
\par
Since the Ly$\alpha$ emission line is narrow (${\rm FWHM} =930$ km
s$^{-1}$) and
there are no broad emission lines such as \ion{C}{4}, \ion{C}{3}], and
\ion{N}{5}
in the rest UV spectrum of SDSS~J0839+3805,
we carried out near-infrared spectroscopy in order to examine properties of
H$\beta$ and H$\alpha$ emission lines.
According to its redshift of 2.3163 which is measured by the peak of Ly$\alpha$
emission line, H$\beta$ and H$\alpha$ are expected to
be redshifted into $H$ and $K$ bands, respectively.
Note that all wavelengths in this paper are vacuum wavelengths.

\section{OBSERVATIONS AND DATA REDUCTION} \label{Obs}
The $K$- and $JH$-band spectra of SDSS~J0839+3805 were obtained with the
Cooled Infrared Spectrograph and Camera for OHS
(CISCO) \citep{Mot02}
attached to the Subaru 8.2-m telescope \citep{Iye04} on 2005 March 26 (UT).
It was cloudy, but the seeing was good ($\sim 0.5$\arcsec~in $K$ band).
The slit width was set to be 0.6\arcsec.
This results in a resolution of 50 \AA~and 46 \AA~in $K$ band and $JH$
band (i.e. $R \sim 300 - 400$), respectively, which were measured
by using night sky lines.
The slit position angle was 0\arcdeg.
We obtained 4 exposures in each band, dithering the telescope to observe the quasar
at two positions with a separation of 10\arcsec~along the slit.
The total integration times on the quasar were 1000~s (250~s $\times 4$)
in $K$ band and 1200~s (300~s $\times 4$) in $JH$ band.
The A0 star SAO~061318 was observed immediately after the observation of SDSS~J0839+3805
for sensitivity calibration and removal of atmospheric absorption lines.
\par
The data were reduced using IRAF\footnote{IRAF is distributed by the
National Optical Astronomy
Observatories, which is operated by the Association of Universities for
Research in Astronomy, Inc. (AURA) under
cooperative agreement with the National Science Foundation.}
in the standard procedure of
flat fielding, sky subtraction, and residual sky subtraction.
Wavelength calibration was performed using OH night sky lines.
The rms wavelength calibration error is 2.6 \AA~in $K$ band and 
1.5 \AA~in $JH$ band, corresponding to 36 km s$^{-1}$ at 21780 \AA~(redshifted
H$\alpha$ line of the SDSS~J0839+3805) and 28 km s$^{-1}$ at 16170 \AA~(redshifted H$\beta$ line), respectively.
The sensitivity calibration was performed as a function of wavelength
and the atmospheric
absorption feature was removed by using the spectrum of SAO~061318.
Since the weather condition was not photometric, the absolute flux
calibration was made using photometry in the 2MASS Point Source 
Catalog as described later.

\section{RESULTS} \label{Results}
Figure \ref{fig2} displays the $K$-band spectrum of SDSS~J0839+3805
as well as the fitting result.
It clearly shows the presence of a strong H$\alpha$ broad emission line with an
absorption  line.
The absorption line is clearly visible in all the spectra from the individual
exposures.
The absorption line must be real since there is no atmospheric
absorption line at this wavelength
and no bad pixel at this position of the detector.
In addition, the spectrum of the standard star does not show any absorption 
features at
the same wavelength.
The H$\alpha$ emission line was fitted with a
combination of three Gaussians and a linear continuum
by using the wavelength regions of the spectra free of absorption features.
The model profile thus constructed is shown as a solid line in Figure \ref{fig2}.
The peak wavelength of H$\alpha$ of the model profile is
$21781\pm4 $\AA, and the redshift is $2.3179\pm0.0006$.
The H$\alpha$ emission therefore appears to be slightly redshifted
($\sim$ 140 km s$^{-1}$) compared to Ly$\alpha$.
However, the uncertainty of the redshift is larger than this
nominal error due to the significant absorption around the
emission line peak; it would be about $\pm 0.002$ (180 km s$^{-1}$).
The H$\alpha$ emission is thus at the same velocity as Ly $\alpha$
emission within the uncertainty.
The FWHM is corrected for the instrumental broadening by using the simple
assumption:
$\rm{FWHM}_{true}=(\rm{FWHM}_{obs}^{2} - \rm{FWHM}_{inst}^{2})^{1/2}$,
where $\rm{FWHM}_{obs}$ is the observed FWHM of the line and
$\rm{FWHM}_{inst}$ is an
instrumental FWHM.
The $\rm{FWHM}_{true}$ of H$\alpha$ is $6500\pm50$ km s$^{-1}$ and is much
broader than that of Ly$\alpha$.
In order to measure the equivalent width (EW) of H$\alpha$ absorption,
the spectrum is normalized by its {\it effective} continuum which includes
both the continuum and broad emission line.
The rest EW of the absorption line is calculated to
be 8.0 {\AA}.
However, since the absorption is strong enough to distort
the emission-line profile, it is hard to accurately restore
the observed emission-line profile to the original one
without the absorption line. This indicates that there is
significant uncertainty in the effective continuum and
therefore the EW of the H$\alpha$ absorption. We estimate a
lower limit of this EW by using an extreme case of effective
continuum, which is shown as the dotted line in Figure 2.
The lower limit of EW thus obtained is 4.9 \AA.
The redshift determined by the bottom of the H$\alpha$ absorption line is
$2.3121\pm0.0002$,
and is blueshifted by 520 km s$^{-1}$ relative to the H$\alpha$ emission
line.
The $\rm{FWHM}_{obs}$ of the absorption line is $770\pm50$ km s$^{-1}$.
Since the $\rm{FWHM}_{inst}$ is $\sim 690$ km s$^{-1}$ at $K$ band,
the line is marginally resolved and
the $\rm{FWHM}_{true}$ is $340^{+100}_{-130}$ km s$^{-1}$.
The properties of absorption lines and emission lines are tabulated in
Table~\ref{tbl2}.
\par
Figure \ref{fig3}a displays the $H$-band spectrum of SDSS~J0839+3805.
The signal-to-noise ratio of the spectrum is not high
because of the worse weather and strong OH sky emission lines.
The feature at $1.617 \mu$m is H$\beta$ emission line,
and the feature at $1.659 \mu$m is probably [\ion{O}{3}] $\lambda5008$
\footnote{This is vacuum wavelength.}.
A narrow absorption-like feature is seen at $1.609 \mu$m in the
H$\beta$ emission line.
However, since its width (25 \AA) is significantly narrower
than the spectral resolution (46 \AA), it is not real. 
The $H$-band spectrum was modeled with a combination of
Gaussians representing the H$\beta$ broad emission line and
[\ion{O}{3}] $\lambda\lambda 4960, 5008$ emission lines, and a linear 
continuum. 
The result is displayed in Figure \ref{fig3}a with a solid line
as well as in Table~\ref{tbl2}.
In order to trace the asymmetric shape, we fitted  the H$\beta$
emission profile with two Gaussians.
We ignored the possible absorption 
feature corresponding to the H$\alpha$ absorption.
The peak of the model H$\beta$ emission-line profile is at $1.617 \mu$m ($z=2.325$) and
it is redshifted by 640 km s$^{-1}$ relative to the H$\alpha$ emission.
This apparent redshift relative to the H$\alpha$ emission is probably 
caused by absorption at the blue side of the H$\beta$ emission line.
The $\rm{FWHM}_{true}$ of H$\beta$ emission is $3280 
\pm 140$ km s$^{-1}$, which is
much narrower than that of H$\alpha$ emission.
This would also be expected if an absorption line exists and
distorts the H$\beta$ emission at the blue side as stated
above.
\par
We fitted [\ion{O}{3}] $\lambda\lambda 4960, 5008$ with a Gaussian for
each line.
The width and redshift were assumed to be the same for both of these
two [\ion{O}{3}] lines (see Table~\ref{tbl2}),
and the intensity ratio of
[\ion{O}{3}] $\lambda 5008$ to $\lambda 4960$ was fixed to be 3.0.
The peak of [\ion{O}{3}] $\lambda 5008$ is at $1.6590\pm0.0005 \mu$m
($z=2.3126\pm0.0009$).
The [\ion{O}{3}] emission line is blueshifted by 480 km s$^{-1}$
relative to the H$\alpha$ emission.
The width of [\ion{O}{3}] emission line is resolved and 
the $\rm{FWHM}_{true}$ is $1310\pm90$ km s$^{-1}$.
Such broad and blueshifted [\ion{O}{3}] emission line has been discovered in
more than one dozen of narrow-line Seyfert1s and narrow-line quasars
\citep{Aok05,mar03,zam02}.
We note that rest-frame EW of [\ion{O}{3}] $\lambda 5008$ is $13 \pm 1$ \AA~for SDSS~J0839+3805,
which is much larger than those found in other LoBALs/FeLoBALs.
All four LoBALs/FeLoBALs in \citet{BM92} have no or weak [\ion{O}{3}] $\lambda 5008$
(EW $< 2$ \AA).
The EWs of six LoBALs in \citet{YW03} are less than 6 \AA.
The current observations suggest that [\ion{O}{3}] in LoBALs/FeLoBALs is 
weak in general, 
and the strength of  [\ion{O}{3}] in SDSS~J0839+3805 is exceptional.
There is, however, only a dozen observations of [\ion{O}{3}] 
in LoBALs/FeLoBALs.
We should wait for more observations.
\par
The $J$-band spectrum is shown in Figure \ref{fig3}b.
The possible absorption feature at $1.288 \mu$m may be 
\ion{He}{1} $\lambda3889.74$.
If the absorption is real, its redshift is $2.3114 \pm 0.0009$, and 
rest EW is $4.8\pm0.7$\AA.
The line is marginally resolved and the $\rm{FWHM}_{true}$ is
$390^{+300}_{-390}$ km s$^{-1}$.
The redshift and FWHM are similar to those of the H$\alpha$ absorption.
\par

\section{DISCUSSION} \label{Discussion}
\subsection{H$\alpha$ Absorption}
Our spectroscopy discovers the presence of an H$\alpha$ absorption
line in SDSS~J0839+3805.
This is the first case for detection of the H$\alpha$ absorption among quasars.
The H$\alpha$ absorption of SDSS~J0839+3805 is deeper than continuum height.
Since an absorber cannot remove more light from a continuum than is initially
present and the depth of the absorption exceeds the height of the continuum,
the absorption line is of non-stellar origin.
We cannot consider the case where only the continuum suffers from absorption.
Two cases are possible:
both the continuum and broad emission-line region suffers from absorption
or only the broad emission-line does.
In either case the absorber is thus located outside
the broad emission-line region, and absorbs at least partially the light from 
the broad emission-line region.
\par
In order to compare the H$\alpha$ absorption to UV absorption lines,
we measured their redshifts and widths in the spectrum of SDSS DR3.
We selected low-ionization lines and less blended lines.
The results are tabulated in Table \ref{tbl3}.
The redshifts of UV low-ionization absorption lines are almost the same as 
that of the H$\alpha$ absorption line.
The widths of the UV low-ionization absorption lines are 360 (\ion{Zn}{2}) -- 880
(\ion{Si}{2}) km s$^{-1}$
which are similar to that of H$\alpha$ absorption.
These facts suggest that the H$\alpha$ and UV low-ionization absorption
lines are produced by the same outflowing low-ionization gas which contains
a substantial amount of neutral gas.
\par
Since our spectroscopy resolved marginally the absorption line profile,
the column density  can be derived from the curve of growth \citep{Sp78}.
The FWHM$_{\rm true}$ of the H$\alpha$ absorption line is 340 km s$^{-1}$,
corresponding to the Doppler parameter of $b = 200$ km s$^{-1}$
[$b=$FWHM/(4 ln 2)$^{1/2}$].
The EW of 8.0 \AA~ is on the ``linear'' regime of the curve of growth 
for H$\alpha$ with $b=$200 km s$^{-1}$,
and the column density  $N_{j}$ for the hydrogen atoms which absorb
H$\alpha$ photons
is $3.2 \times 10^{13} {\rm cm}^{-2}$.
In the case of the extremely low EW (4.9 \AA), the column density would be
$2.0 \times 10^{13} {\rm cm}^{-2}$.
Assuming that the absorber is in thermal equilibrium at a
temperature of 10,000 K, the number ratio of hydrogen atoms
at the $n=2$ level relative to those at the ground level is
calculated to be $3.0 \times 10^{-5}$ using the Boltzmann equation.
Therefore the neutral hydrogen column density is
expected to be $\sim 10^{18} {\rm cm}^{-2}$. 
The lack of information about the temperature of the
absorber is the limitation of our analysis; high resolution
spectroscopy of the H $\alpha$ absorption is needed. 
We cannot be certain whether the absorption is saturated or not from our low-resolution spectroscopy.
As a result, we cannot determine whether the absorber fully covers the continuum and 
the broad emission-line region.
In case of the partial coverage of either sources,
our {\it effective} continuum used in the above analysis would be an overestimate. 
The EW and column density would therefore be underestimates.
On the other hand, since the level population is very sensitive to a temperature,
the number ratio of hydrogen atoms
at the $n=2$ level relative to those at the ground level would be larger
if the temperature is higher than 10,000 K.
The column density would be overestimate in that case. 
The uncertainty in temperature may cause an orders-of-magnitude error in 
column density.

\subsection{Reddening of the Continuum} \label{reddening}
Since the color of SDSS~J0839+3805 
($g-r=0.698$ mag and $g-i=0.678$ mag) is redder
than the median color of quasars at $z=2.3$ 
($g-r=0.08$ mag and $g-i=0.2$ mag) in SDSS \citep{Ric03},  
the spectrum is probably reddened.
In order to estimate the reddening,
we assume the presence of dust in the quasar with the
SMC-like reddening law \citep{Pei92}.
We also tried two other types of reddening law (Milky Way and LMC),
but the Milky Way-like reddening law is rejected by the absence of 
the 2200 \AA~feature in the spectrum.
Though LMC-like reddening law is acceptable, the SMC-like reddening law gives us a better fit to the observed spectrum.
We also assume the composite quasar spectrum from the SDSS \citep{VB01} as
the unreddened spectrum.
The spectral energy distribution (SED) of SDSS~J0839+3805 is composed of 
the spectrum from the SDSS DR3, the photometry from 2MASS
and our spectra, and shown in Figure {\ref{fig4}.
We also plotted the SDSS photometric data except for $u$ band with the SED in Figure {\ref{fig4}.
Our spectra were scaled to the 2MASS photometric data.
Since our $J$- and $H$-band spectra were taken simultaneously, the scaling
factors should be the same value.
But they are different by 50\%.
As seen in Figure \ref{fig4},
the $J$-band 2MASS photometry looks discrepant from the SDSS data,
2MASS $H$- and $K$-band data.
We therefore scaled both $J$- and $H$-band spectra by using the 
factor adequate for the $H$-band spectrum.
The SDSS photometric data match the SDSS spectrum well.
\par
The SED of SDSS~J0839+3805 is reasonably well reproduced by a reddened
SDSS composite spectrum with a reddening of $E(\bv) \sim 0.15$ mag,
which is shown by a blue dashed line in Figure \ref{fig4};
the reddened composite spectrum delineates the continuum level of
the spectrum of SDSS~J0839+3805.
This amount of reddening is larger than those derived from comparisons
between the composite spectrum of LoBALs and that of non-BAL quasars.
\citet{Bro01} estimated that LoBALs composite is reddened by
 $E(\bv) \sim 0.1$ mag by using the data from FBQS, and
\citet{Rei03} found the average reddening for LoBALs is $E(\bv) \sim
0.08$ mag from SDSS data.
However, two known FeLoBALs are extremely reddened.
Hawaii 167 has $E(\bv) \simeq 0.54 - 0.7$ mag \citep{Egami96}, and
FIRST~J155633.8+351758 has  $E(\bv) \simeq 0.6$ mag \citep{Naji00}.
\citet{Hal02} discovered several FeLoBALs reddened by
$E(\bv) \sim 0.1 - 0.7$ mag. 
FeLoBALs may be more reddened than LoBALs.
\citet{Hop04} also found 2\% of quasars are reddened by $E(\bv) > 0.1$ mag
by examining the observed distribution of $z < 2.2$ quasar SEDs
including BAL quasars
in the SDSS first data release quasar catalog \citep{Sch03}.
\par

\subsection{Similarity to NGC 4151} 
To date, the presence of non-stellar Balmer absorption lines in an active galactic nucleus
(AGN) is known only in spectrum of NGC 4151 \citep{AK69, Ser99, Hut02}.
The EW of H$\alpha$ absorption in NGC 4151 was found to be $2.33 - 5.17$ \AA~
\citep{And74, Ser99},
and variable \citep{And74, Hut02}.
The kinematical characteristics of Balmer absorption are similar
to those of H$\alpha$ absorption in SDSS~J0839+3805.
The velocity blueshift of the H$\beta$ absorption was $\sim 500$ km s$^{-1}$
and FWHM were $340 - 420$ km s$^{-1}$ in NGC 4151,
when the continuum flux of the nucleus was low and the absorption was deep
\citep{Hut02}.
Interestingly, when the continuum flux of the nucleus was low,
there were a large number of \ion{Fe}{2} absorption lines that arise
from excited fine-structure
levels of the ground term and excited terms in the UV spectrum of NGC 4151
\citep{Cre00, Kra01}.
\par
We show the UV spectrum of NGC 4151 in the low state from
\citet{Kra01}, and compare it to that of SDSS~J0839+3805 in Figures \ref{fig5}, \ref{fig6}, and \ref{fig7}.
The low-state UV spectrum of NGC 4151 is very similar to that of
SDSS~J0839+3805, particularly absorption features
seen in the wavelength region between 2000 \AA~and 2700 \AA,
though the features of the emission lines such as \ion{N}{5},
\ion{C}{4}, and \ion{C}{3}] are different.
The remarkable similarity to the low-state UV spectrum of NGC 4151
implies that
the conditions of absorbing gas in SDSS~J0839+3805 are similar
to those in NGC 4151,
though a high-resolution spectrum as well as a detailed modeling of it
are required to derive a firm conclusion.
We also note that rare \ion{He}{1} $\lambda3890$ absorption line
is found in both objects \citep{AK69,And74} although
it is a possible detection in SDSS~J0839+3805.
Only less than 10 previously known AGNs show \ion{He}{1} $\lambda3890$
absorption line \citep{Hal02}.
\par

From the morphology of narrow-line region, NGC 4151 is thought to be
looked at the direction close to the edge of the obscuring torus
\citep{Eva93}.
SDSS~J0839+3805 may be also a quasar seeing from the edge of the torus
\citep{GM95, Cohen95, HW95, SH99}.
As described in \S\ref{reddening}, SDSS~J0839+3805 is indeed
reddened by $E(\bv) \sim 0.15$ mag,
although the dust enshrouded quasar
model cannot be excluded.

\section{CONCLUSION} \label{Conclusion}
We discovered an H$\alpha$ absorption in the unusual BAL quasar 
SDSS~J0839+3805 with near-infrared spectroscopy.
This is the first case for detection of H$\alpha$ absorption in quasars.
The H$\alpha$ absorption in SDSS~J0839+3805 is blueshifted by 520 km
s$^{-1}$
relative to H$\alpha$ emission line, and its redshift is almost coincident
with those of UV low-ionization metal absorption lines.
The width of H$\alpha$ absorption line is similar to those of
the UV low-ionization metal absorption lines ($\sim 340$ km s$^{-1}$).
These facts suggest that H$\alpha$ and low-ionization metal absorption lines
are produced by the same low-ionization gas
which has a substantial mount of neutral gas.
The column density of the neutral hydrogen is expected to be $10^{18}$
cm$^{-2}$ from the analysis of the curve of growth
by assuming that the absorbing gas is in thermal equilibrium at
a temperature of 10,000 K.
SDSS~J0839+3805 is found to have $E(\bv) \sim 0.15$ mag
when the SED is compared to the quasar composite spectrum.
The similarity of UV spectrum of SDSS~J0839+3805 to
that of NGC 4151 in its low state is remarkable.
This fact suggests the physical conditions of the absorber in 
SDSS~J0839+3805 are similar to those of NGC 4151 in the low state.
As proposed for NGC 4151, SDSS~J0839+3805 is also considered to be seen through
the close direction of the surface of the obscuring torus.
\par
Observations of H$\beta$ with a much better signal-to-noise-ratio are necessary
to measure the EW of H$\beta$ absorption.
Observations of the H$\alpha$ absorption line with a higher spectral
resolution will reveal whether the absorption is a saturated line and 
whether the absorber fully covers the continuum and the broad emission-line
region. 
These observations will enable us to estimate the column density of neutral
hydrogen more accurately .
High resolution spectroscopy in the optical domain (i.e., rest UV) will also
allow us to estimate the column densities of many ions.
The ratios of absorption lines from the excited level to that from the
ground level,
such as \ion{Si}{2*} $\lambda1533$\footnote{The absorption arising from
excited fine-structure levels of the ground term of \ion{Si}{2} is
denoted as \ion{Si}{2*}.}/\ion{Si}{2} $\lambda1527$, provide direct
measurements of the electron density. 
The ionization degree and density of absorbing gas 
will help us to locate the absorbing gas and know the geometry of the 
absorber around the nucleus.
There are one dozen H$\alpha$ observations of FeLoBALs
\citep[][K. Aoki et al., in preparation]{BM92,Egami96,Lacy02,Naji00,Brun03}.
To date, no H$\alpha$ absorption line has been found except for
SDSS~J0839+3805.
This may be happened due to such variability as found in NGC 4151 \citep{Cre00}.
The physical origin of H$\alpha$ absorption is an open question.
More spectroscopic data of FeLoBALs in the near-infrared would answer it.

\acknowledgments
We are grateful to the staffs of Subaru Telescope for their assistance
during our observations.
We are also grateful to Ryuko Hirata, Bev Wills and Kozo Sadakane
for useful discussions and comments.
We also thank the referee for detailed comments which improved the paper. 
KO's activity is supported by a Grant-in-Aid
for Scientific Research from Japan Society for
Promotion of Science (17540216).
Funding for the Sloan Digital Sky Survey (SDSS) has been provided by the
Alfred P. Sloan
Foundation, the Participating Institutions, the National Aeronautics and
Space Administration,
the National Science Foundation, the U.S. Department of Energy, the
Japanese Monbukagakusho, and
the Max Planck Society. The SDSS Web site is \url{http://www.sdss.org/}.
The SDSS is managed by the Astrophysical Research Consortium (ARC) for
the Participating Institutions.
The Participating Institutions are The University of Chicago, Fermilab,
the Institute for Advanced Study,
the Japan Participation Group, The Johns Hopkins University, the Korean
Scientist Group,
Los Alamos National Laboratory, the Max-Planck-Institute for Astronomy
(MPIA),
the Max-Planck-Institute for Astrophysics (MPA), New Mexico State
University, University of Pittsburgh,
University of Portsmouth, Princeton University, the United States Naval
Observatory, and the
University of Washington.
This publication makes use of data products from the Two Micron All Sky
Survey, which is a joint project of the University of Massachusetts and
the Infrared Processing and Analysis Center/California Institute of
Technology, funded by the National Aeronautics and Space Administration
and the National Science Foundation.

\clearpage

\begin{figure}
\plotone{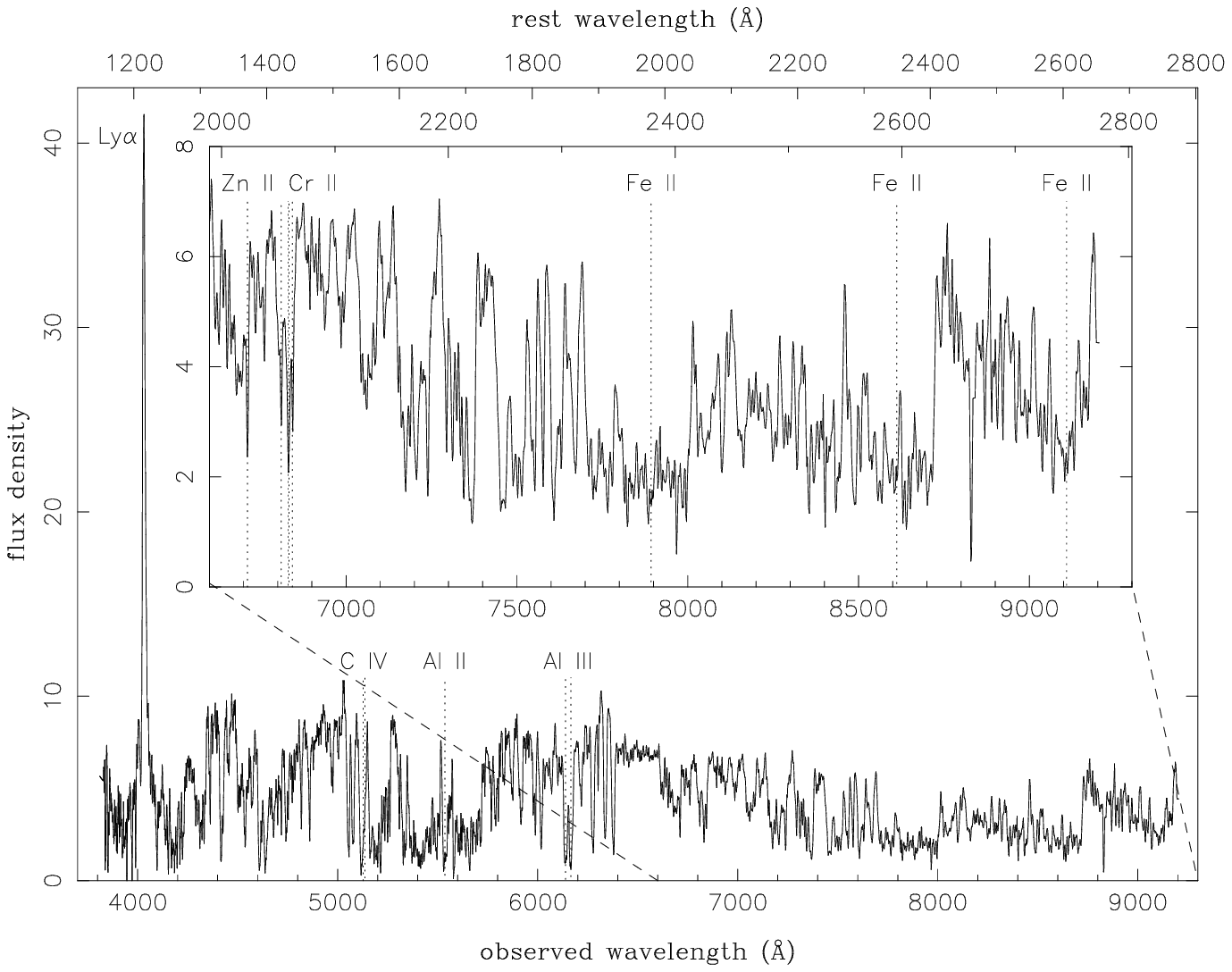}
\caption{Observed spectrum of SDSS~J0839+3805 in the SDSS DR3.
Ordinate is a flux density in units of $10^{-17}$ erg s$^{-1}$ cm$^{-2}$
\AA$^{-1}$
and abscissa is an observed wavelength in angstrom.
The rest wavelength is given along the top axis.
The spectrum is smoothed by a 3-pixel boxcar filter.
Dotted lines show the wavelengths of the absorption lines.
\label{fig1}}
\end{figure}
\clearpage

\begin{figure}
\plotone{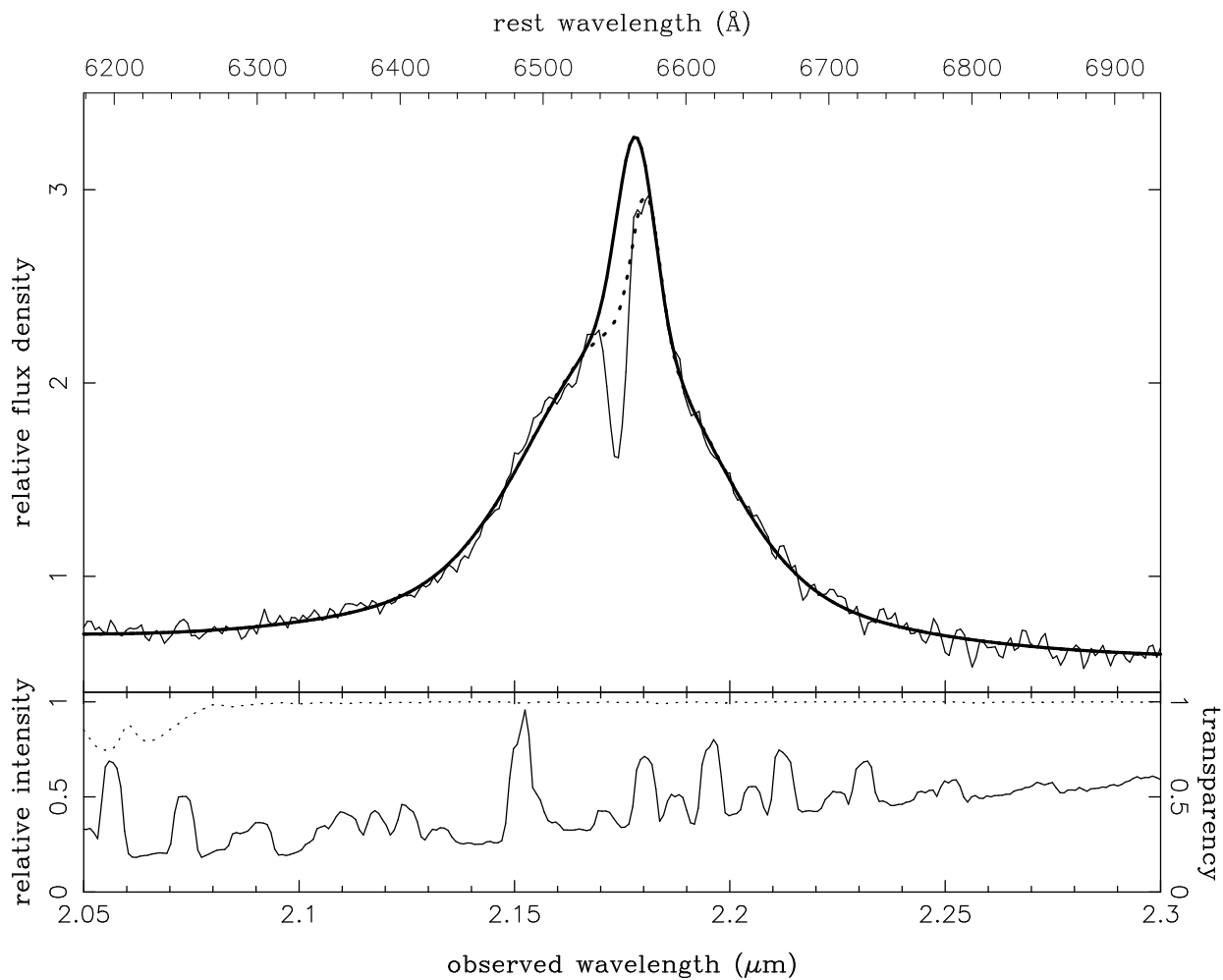}
\caption{(Upper panel:) $K$-band spectrum of SDSS~J0839+3805.
Ordinate is a relative flux density in units of erg s$^{-1}$ cm$^{-2}$
$\mu$m$^{-1}$
and abscissa is an observed wavelength in vacuum in micron.
The rest wavelength is given along the top axis.
The H$\alpha$ emission is fitted with three Gaussians.
The best fit is shown as a thick solid line.
The extreme case of the effective continuum is shown by a
dotted line.
(Lower panel:) The sky emission spectrum (solid line) and the
atmospheric transmission curve
(dotted line) which is obtained from
the UKIRT WWW page.
It is produced using the program IRTRANS4.
\label{fig2}}
\end{figure}
\clearpage

\begin{figure}
\epsscale{0.6}
\plotone{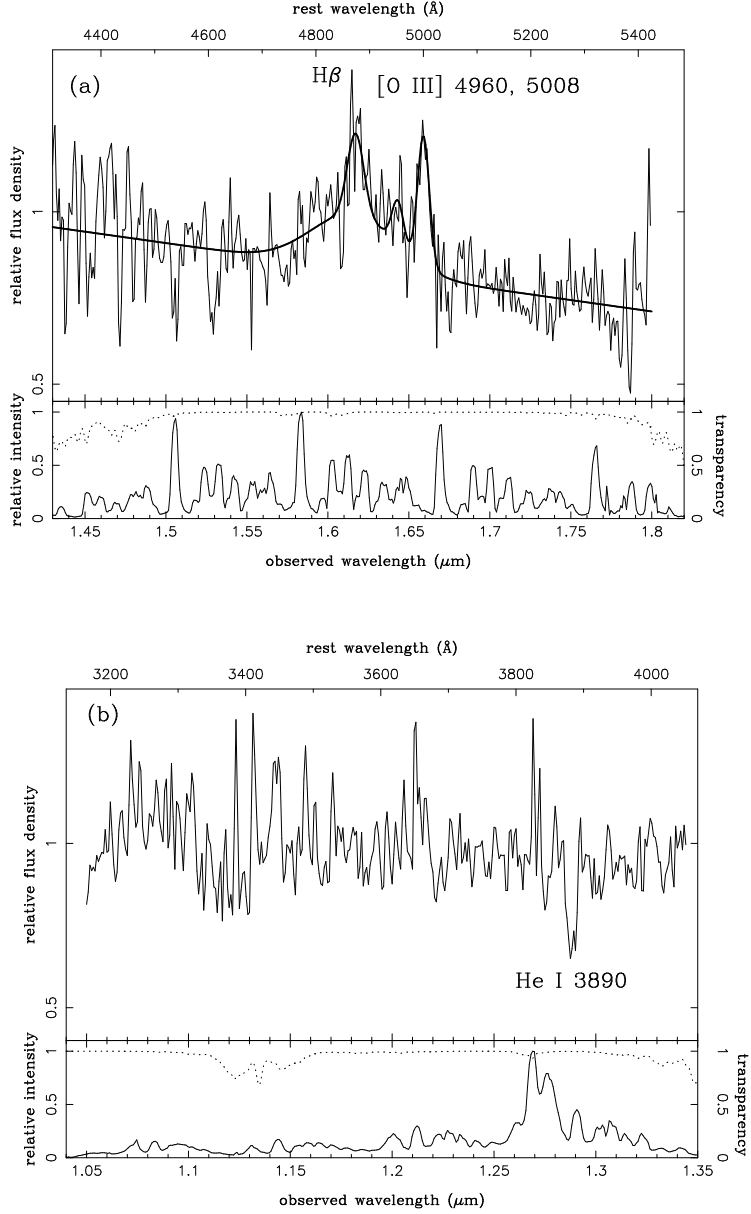}
\caption{(a)(Upper panel:) $H$-band spectrum of SDSS~J0839+3805.
Ordinate is a relative flux density in units of erg s$^{-1}$ cm$^{-2}$
$\mu$m$^{-1}$
and abscissa is an observed wavelength in vacuum in micron.
The rest wavelength based on the redshift of H$\alpha$ emission line 
is given along the top axis.
The H$\beta$ emission is fitted with two Gaussians, and 
[\ion{O}{3}] $\lambda\lambda 4960, 5008$ are fitted with fixed intensity ratio 
of 3.0 and the same redshift and width of a Gaussian for each line.
(Lower panel:) Same as Fig. 2.
(b) (Upper panel:) $J$-band spectrum of SDSS~J0839+3805.
Ordinate is a relative flux density in units of erg s$^{-1}$ cm$^{-2}$
$\mu$m$^{-1}$
and abscissa is an observed wavelength in vacuum in micron.
The rest wavelength is given along the top axis.
(Lower panel:) Same as Fig. 2.
\label{fig3}}
\end{figure}
\clearpage

\begin{figure}
\epsscale{1.0}
\plotone{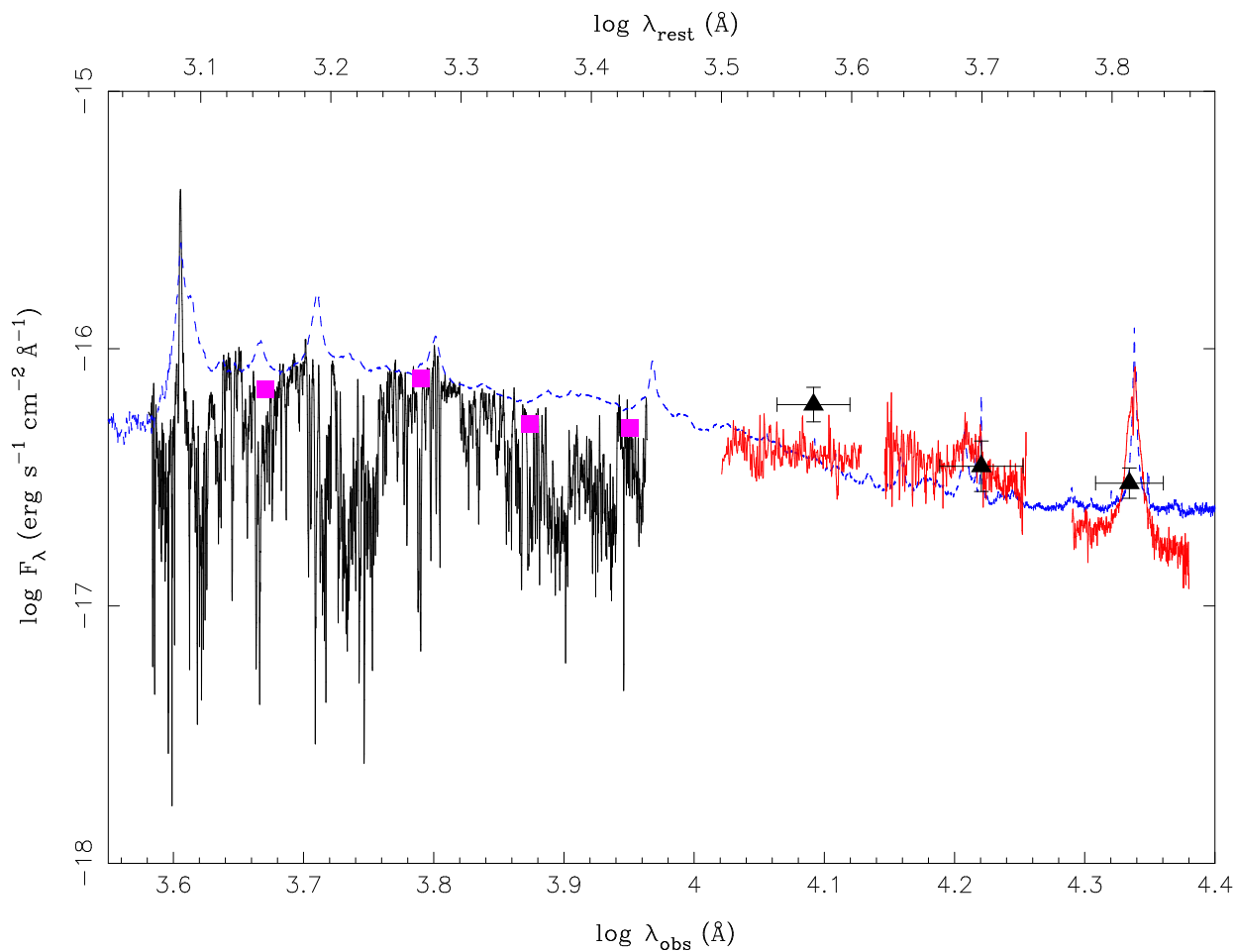}
\caption{Spectral energy distribution of SDSS~J0839+3805.
Ordinate is logarithm of an observed flux density in units of erg
s$^{-1}$ cm$^{-2}$ \AA$^{-1}$
and abscissa is logarithm of an observed wavelength in vacuum in angstrom.
The rest wavelength is given along the top axis.
The rest UV spectrum by the SDSS and rest optical spectra by us are
shown with a black line and red lines,
respectively.
Our rest optical spectra are scaled to the photometry by 2MASS (black
triangles),
except for $J$ band (see text).
The SDSS photometric data are shown in magenta squares.
The SDSS composite quasar spectrum reddened by the SMC-type extinction law
with  $E(\bv)=0.15$ mag is shown with a blue dashed line.
\label{fig4}}
\end{figure}
\clearpage

\begin{figure}
\epsscale{0.8}
\plotone{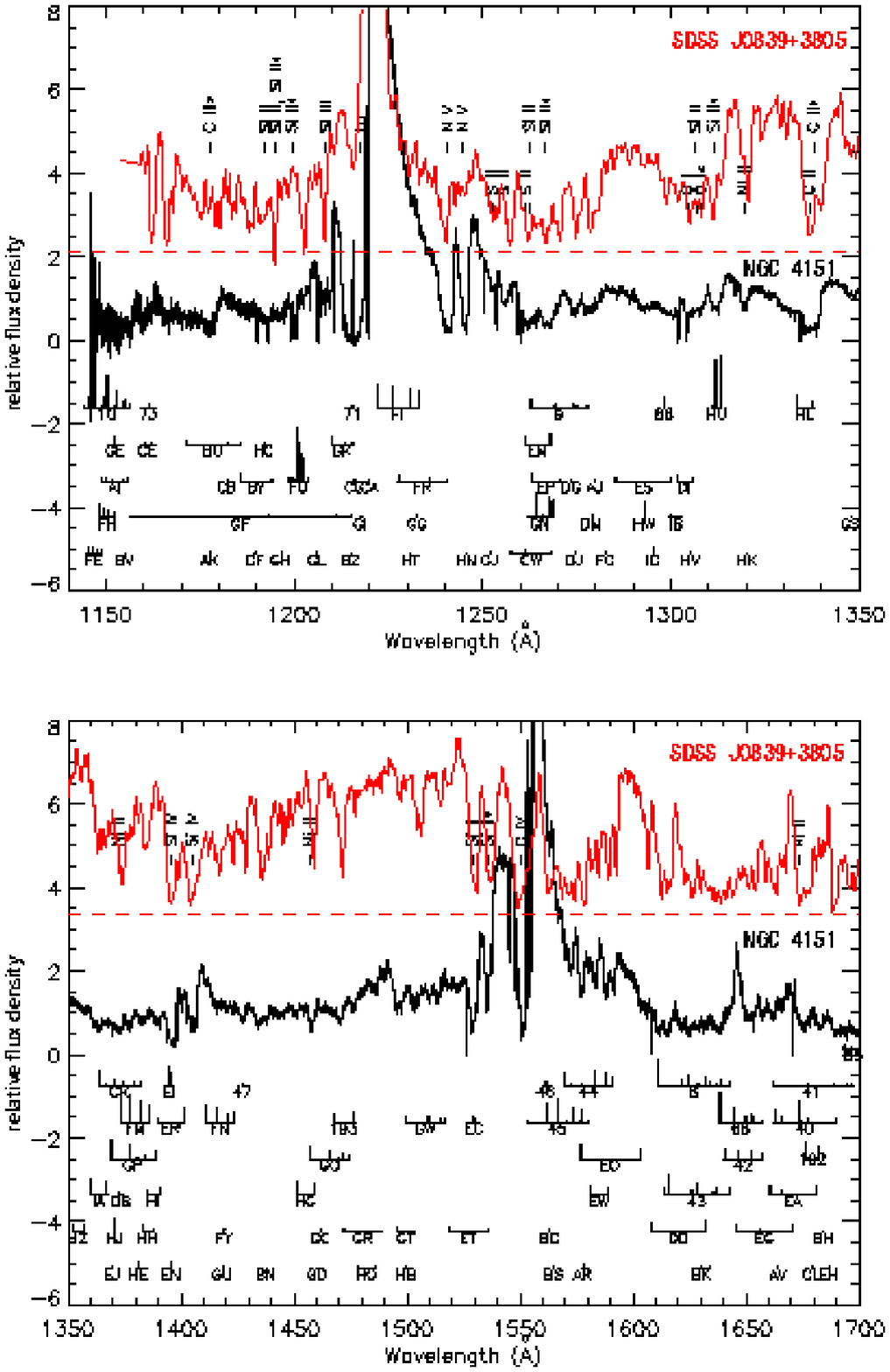}
\caption{Comparison of UV spectrum of SDSS~J0839+3805 (upper) with that of
NGC 4151 (lower) in the low state taken from \citet{Kra01}.
The spectrum of SDSS~J0839+3805 is shown in a red line, and shifted to
the observed frame of NGC~4151 in order to compare the absorption features
easily.
Ordinate is a relative flux density in units of erg s$^{-1}$ cm$^{-2}$
\AA$^{-1}$.
The spectrum of SDSS~J0839+3805 is shifted by arbitrary value for clarity.
The zero level of the spectrum of SDSS~J0839+3805 is shown in a red
dashed line.
\ion{Fe}{2} multiplets are plotted below the spectra \citep[see][]{Kra01}.
\label{fig5}}
\end{figure}
\clearpage

\begin{figure}
\epsscale{0.7}
\plotone{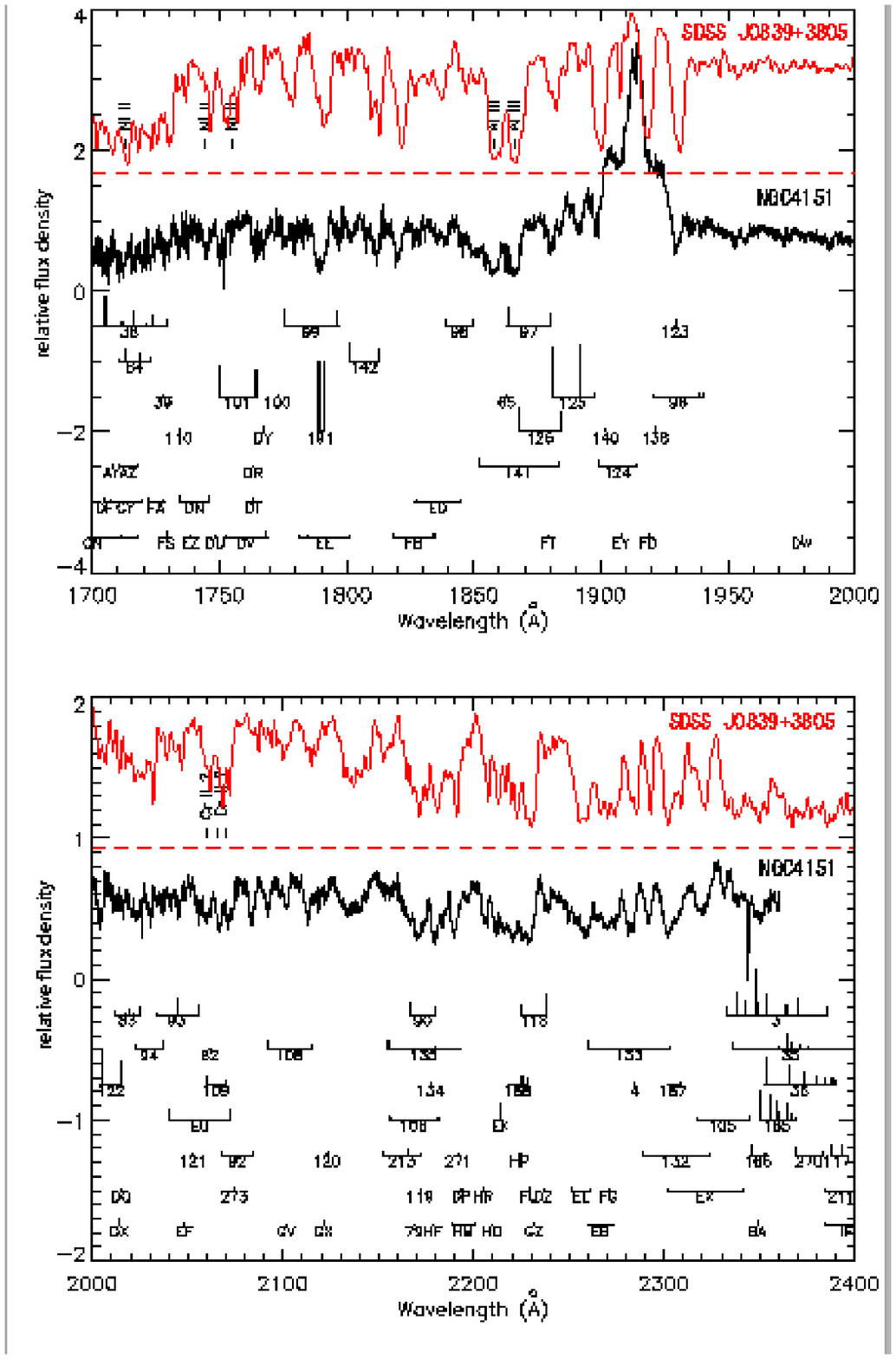}
\caption{Same as Fig. 5, but for the region between 1700 \AA~ and 2400 \AA.
\label{fig6}}
\end{figure}
\clearpage

\begin{figure}
\epsscale{0.7}
\plotone{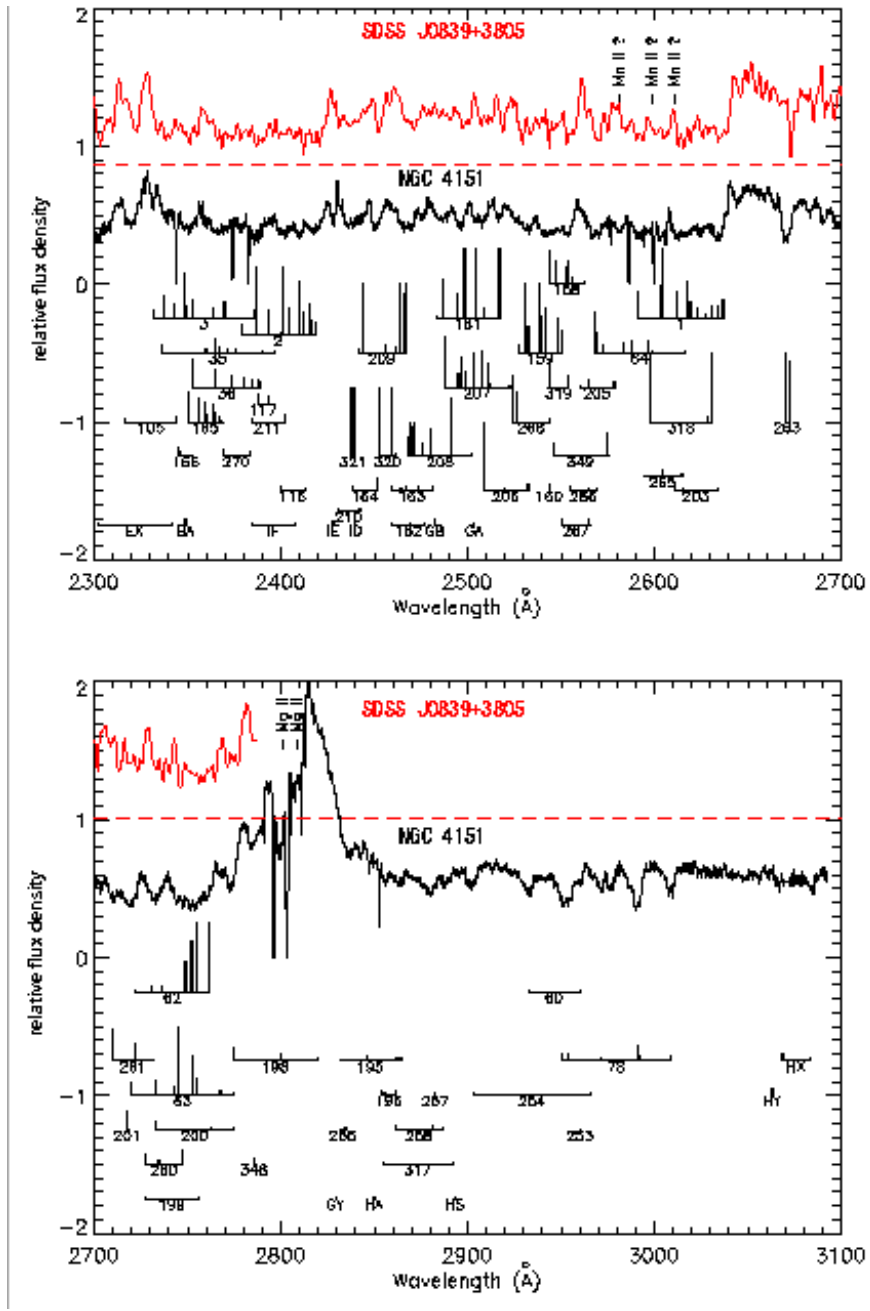}
\caption{Same as Fig. 5, but for the region between 2300 \AA~ and 3100 \AA.
\label{fig7}}
\end{figure}
\clearpage


\begin{deluxetable}{cc}
\tablecaption{Photometric data of SDSS~J0839+3805 \label{tbl1}}
\tablewidth{0pt}
\tablehead{
\colhead{Band} & 
\colhead{Magnitude} \\
}
\startdata
$u$  & $21.597\pm0.149$\tablenotemark{a}\\
$g$  & $19.631\pm0.020$\tablenotemark{a}\\
$r$  & $18.933\pm0.022$\tablenotemark{a}\\
$i$  & $18.953\pm0.019$\tablenotemark{a}\\
$z$  & $18.604\pm0.034$\tablenotemark{a}\\
$J$  & $16.782\pm0.167$\tablenotemark{b}\\
$H$  & $16.278\pm0.245$\tablenotemark{b}\\
$Ks$ & $15.386\pm0.147$\tablenotemark{b}\\
\enddata
\tablenotetext{a}{Magnitude measured by PSF-fitting
          photometry in AB system.}
\tablenotetext{b}{Magnitude in Vega system.}
\end{deluxetable}

\begin{deluxetable}{ccccc}
\tablecaption{Properties of Absorption and Emission Lines \label{tbl2}}
\tablewidth{0pt}
\tablehead{
\colhead{Line} &
\colhead{Absorption/Emission} & 
\colhead{$z$} &
\colhead{$\Delta v$\tablenotemark{a}} & 
\colhead{$\rm{FWHM}_{true}$}\\
\colhead{} & 
\colhead{} & 
\colhead{} &
\colhead{(km s$^{-1}$)}& 
\colhead{(km s$^{-1}$)}
}
\startdata
H$\alpha$ & Emission   & $2.3179\pm0.002$ & \nodata  & $6500\pm50$ \\
H$\alpha$ & Absorption & $2.3121\pm0.0002$ & $-520\pm20$  &
$340^{+100}_{-130}$ \\
H$\beta$ & Emission   & $2.325\pm0.002$\tablenotemark{b}
&$+640\pm180$\tablenotemark{b}   & $3280\pm140$\tablenotemark{b} \\
{[\ion{O}{3}]} 5008 & Emission & $2.3126\pm0.0009$ & $-480\pm80$ &
$1310\pm90$ \\
\ion{He}{1} 3890 & Absorption &  $2.3114\pm0.0009$ & $-590\pm80$ &
$390^{+300}_{-390}$ \\
Ly$\alpha$ & Emission   & $2.3163\pm0.0002$ & $-140\pm20$  & $930\pm25$ \\
\enddata
\tablenotetext{a}{Velocity shift relative to the H$\alpha$ emission line.
Negative value means blueshift.}
\tablenotetext{b}{Redshift and FWHM of the H$\beta$ emission line are
probably affected by the absorption.}
\end{deluxetable}

\begin{deluxetable}{cccc}
\tablecaption{Properties of Low-ionization UV Absorption Lines \label{tbl3}}
\tablewidth{0pt}
\tablehead{
\colhead{Line} & 
\colhead{$z$} &
\colhead{$\Delta v$\tablenotemark{a}} &
\colhead{$\rm{FWHM}_{true}$}\\
\colhead{} & 
\colhead{} & 
\colhead{(km s$^{-1}$)} & 
\colhead{(km s$^{-1}$)}
}
\startdata
\ion{Si}{2} 1527 & 2.3100 & $-$720 & 810 \\
\ion{Si}{2} 1533 & 2.3110 & $-$620 & 880 \\
\ion{Al}{3} 1855 & 2.3101 & $-$710 & 890 \\
\ion{Al}{3} 1863 & 2.3101 & $-$710 & 970 \\
\ion{Fe}{3} 1914 & 2.3109 & $-$630 & 730 \\
\ion{Zn}{2} 2026 & 2.3124 & $-$500 & 360 \\
\enddata
\tablenotetext{a}{Velocity shift relative to the H$\alpha$ emission line
($z=2.3179$).
Negative value means blueshift.}
\end{deluxetable}

\end{document}